\documentclass[11pt,letterpaper]{article}
\pdfoutput=1

\usepackage{graphicx,array,multicol}
\usepackage[dvipsnames,table,xcdraw]{xcolor}
\definecolor{darkblue}{rgb}{0,0.1,0.5}
\definecolor{darkgreen}{rgb}{0,0.5,0.2}
\definecolor{darkred}{RGB}{153,26,0}
\usepackage{ulem}
\definecolor{seablue}{rgb}{0,0.2,0.6}
\usepackage{latexsym}
\usepackage{amssymb, amsmath}
\usepackage{slashed}
\usepackage{bm}        % for math
\usepackage[numbers,sort&compress]{natbib}
\usepackage{bm,relsize}
\usepackage{slashed}
\usepackage{tcolorbox}
\definecolor{viola}{RGB}{134,41,198}
\usepackage{mathrsfs}
\usepackage{hyperref} %Automatically links \label and \ref commands; Always load last
\hypersetup{
    colorlinks=true,       % false: boxed links; true: colored links
    linkcolor=darkblue,          % color of internal links
    citecolor=BrickRed,        % color of links to bibliography
    filecolor=magenta,      % color of file links
    urlcolor=MidnightBlue           % color of external links
}
\usepackage[all]{hypcap} %Link navagates to top of figure instead of caption (below fig)
\usepackage{subcaption}

\usepackage[font=small,labelfont=bf,labelsep=period]{caption}

\newcommand{\eq}[1]{~{\rm (\ref{eq:#1})}}
\makeatletter
\font\ital=cmu10

\def\hhref#1{\href{http://arxiv.org/abs/#1}{arXiv:#1}}
\usepackage{xstring}
\newcommand{\hhrefq}[1]{\IfSubStr{#1}{:}{\href{http://inspirehep.net/search?ln=en&ln=en&p=#1&of=hb&action_search=Search&sf=&so=d&rm=&rg=25&sc=0}{InSpire:#1}}{\hhref{#1}}}

\def\art{\@ifnextchar[{\eart}{\oart}}
\def\eart[#1]#2#3#4#5#6{{\rm #2}, {\em #3 \bf #4} {\rm (#6) #5} ({\em #1})}
\def\article{\@ifnextchar[{\earticle}{\oarticle}}
\def\oarticle#1#2#3#4#5#6{{\rm #1}, {\ital `#6'}, {\rm #2 #3 (#5) #4}}
\def\earticle[#1]#2#3#4#5#6#7{{\rm #2}, {\ital `#7'}, {\rm #3 #4 (#6) #5}  [\hhrefq{#1}]}
\def\hepart[#1]#2{{\rm #2, \sl#1}}
\def\heparticle[#1]#2#3{#2, {\ital `#3'} [\hhrefq{#1}]}
\newcommand{\doi}[1]{\href{http://dx.doi.org/#1}{[link]}}

\newcommand{\hhrefqq}[1]{\IfBeginWith{#1}{10.}{\href{https://doi.org/#1}{doi:#1}}{\hhrefq{#1}}}

\renewenvironment{thebibliography}[1]
{\begin{multicols}{2}[\section*{\refname}]%
		\@mkboth{\MakeUppercase\refname}{\MakeUppercase\refname}%
		\list{\@biblabel{\@arabic\c@enumiv}}%
		{\settowidth\labelwidth{\@biblabel{#1}}%
			\leftmargin\labelwidth
			\advance\leftmargin\labelsep
			\@openbib@code
			\usecounter{enumiv}%
			\let\p@enumiv\@empty
			\renewcommand\theenumiv{\@arabic\c@enumiv}}%
		\sloppy
		\clubpenalty4000
		\@clubpenalty \clubpenalty
		\widowpenalty4000%
		\sfcode`\.\@m}
	{\renewcommand{\@noitemerr}
		{\@latex@warning{Empty `thebibliography' environment}}%
		\endlist\end{multicols}}

\usepackage{stackengine}
\stackMath
\newcommand{\be}{\begin{equation}}
\newcommand{\ee}{\end{equation}}

\newcommand{\Op}{\mathcal{O}}

\definecolor{lightergray}{rgb}{0.9,0.9,0.9}

\DeclareMathOperator{\tr}{tr}

\makeatletter
\g@addto@macro\bfseries{\boldmath}
\makeatother

\begin{document}\thispagestyle{empty}
~
\vspace{-8ex}
\begin{center}
{\LARGE \bf \color{red!25!black}
The Too Visible QCD Axion}\\
\bigskip\vspace{2ex}
{
\large Luca Di Luzio$^a$, Michele Redi$^b$, Alessandro Strumia$^c$, Andrea Tesi$^b$, Arsenii V. Titov$^{d,a}$}
\\[1ex]
{\it \small
$^a$INFN Sezione di Padova,  Padova, Italy \\
$^b$INFN Sezione di Firenze,  Sesto Fiorentino, Italy 
%\\ Department of Physics and Astronomy, University of Florence, Italy 
\\
$^c$Dipartimento di Fisica, Universit\`a di Pisa, Italy \\
$^d$Dipartimento di Fisica e Astronomia, Universit\`a di Padova,  Italy 
}
\end{center}

\vspace{1ex}

\centerline{\bf\large Abstract} 
\begin{quote}\large
Murayama \cite{Murayama:2026ioh} proposed a GeV-scale axion theory where
the up-quark mass term is generated dynamically by the QCD chiral condensate, 
spontaneously breaking a Peccei-Quinn symmetry.
It predicts a too large mass splitting between neutral and charged pions. 
Trying to solve this problem we explore extensions.
Despite some partial improvements, we identify a structural obstruction: 
the new Peccei-Quinn spurion breaks the accidental isospin symmetry of the chiral Lagrangian, leading to an enhanced higher-order operator.
As a consequence, pion scatterings too are distorted. 
We also examine the limit where the axion becomes light, finding that it is excluded by fifth-force constraints.
\end{quote}

% Models where the down quark is also charged reduce the tension but they are still problematic. One choice fixes the $\pi^0$ mass, but has problems with unitarity and stable domain walls. Even allowing for multiple PQ scalars we find similar negative results. 

{\small\tableofcontents}

\vfill

\noindent\line(1,0){188}
{\scriptsize{ \\ E-mail:
\href{mailto:luca.diluzio@pd.infn.it}{luca.diluzio@pd.infn.it}, 
\href{mailto:michele.redi@fi.infn.it}{michele.redi@fi.infn.it}, 
\href{mailto:alessandro.strumia@unipi.it}{alessandro.strumia@unipi.it},
\href{mailto:andrea.tesi@fi.infn.it}{andrea.tesi@fi.infn.it},
\href{mailto:arsenii.titov@unipd.it}{arsenii.titov@unipd.it}.
}
}

\section{Introduction}
In a recent paper Murayama proposed a novel axion theory to address the strong CP problem~\cite{Murayama:2026ioh}, where all the relevant dynamics takes place  at the QCD scale. 
As well known, QCD solves the strong CP problem if the bare mass of the up quark vanishes, $m_u=0$. In this case the $\eta'$ acts as an axion, dynamically relaxing the $\theta$-angle to zero. However, this possibility is  excluded by comparing lattice computations with data \cite{PDG}. 
Building on the massless up quark solution, 
Murayama added a new complex scalar Standard Model (SM) singlet 
 $\varphi$ coupled to up quarks 
with unit Peccei-Quinn (PQ) charge under the rotation of $u_R$.
When quarks develop their condensate, 
$\varphi$ acquires a PQ-breaking vacuum expectation value
effectively giving a mass to the up quark.
The main issue of the proposal is that the $\pi^0$ mass is modified 
\begin{equation}
\frac{\Delta m_{\pi^0}}{m_{\pi^0}}\approx- \frac {m_u}{2(m_u+m_d)}\approx -0.15.
\label{eq:splitting1}
\end{equation}
The observed pion mass splitting $m_{\pi^\pm} - m_{\pi^0}\approx 0.03 m_{\pi^0}$  arises predominantly from electromagnetic self-energy effects, 
with only a small correction from the up-down quark mass difference.
Since lattice computations claim a per-cent level precision on the electromagnetic effect~\cite{1604.07112,2108.05311,2202.11970}, this model is excluded unless the correction to the neutral pion mass can be almost completely canceled.

\medskip

In this work we try to improve this conclusion.
In section~\ref{sec:model} we review and extend the model to general PQ charges. In section~\ref{sec:QCDscaleAxion} we discuss models that could improve the mass splitting, but only with partial success. Neither the Kaplan-Manohar ambiguity as proposed in \cite{Murayama:2026ioh}, nor one-loop corrections to pion masses seem to avoid the problem. The pion mass splitting is only mildly reduced by assuming that the down quark $d_R$ has also  PQ charge. In view of $m_d/m_u \approx 2$, the pion mass splitting could be avoided by assuming PQ charges $n_u/n_d = 2$;  however the new physics gets non-perturbative below the QCD scale in that case. This could be cured by adding new light PQ scalars, but scans in the larger parameter space do not find an allowed solution. Isospin invariance is recovered in a model where the  $s$ quark only is PQ charged, but the new PQ scalars turn out to be too light.

\medskip

In section~\ref{sec:EFT} we identify a structural obstacle to build viable models. Like quark masses, the couplings of $\varphi$ break the accidental SU(3)$_{u,d,s}$ flavour symmetry of massless QCD.
As a result, integrating out QCD and $\varphi$, gives an effective chiral Lagrangian for mesons $U$ that differs from the usual QCD chiral Lagrangian. The operator related to quark masses is replaced by a different higher-order operator of comparable size,
\begin{equation}
\Delta\mathscr{L}_{\rm PQ}\sim
\tr[I_{\rm PQ}^\dag U]\tr[I_{\rm PQ} U^\dag]
\end{equation}
containing the new PQ spurion $I_{\rm PQ}$,
given by $I_{\rm PQ} \propto \mathrm{diag}(m_u\,,0\,,0)$ in the model of~\cite{Murayama:2026ioh}. This spurion breaks the accidental SU(2)$_{u,d}$ isospin symmetry of the leading order chiral Lagrangian, so pion masses get the order one correction of eq.\eq{splitting1}, and pion scatterings too are similarly affected. The general problem of the new theories is that they unavoidably predict order one corrections to meson physics, but the  QCD chiral Lagrangian has been more precisely tested.

In the final section~\ref{sec:invax} we reconsider the  new theory  
abandoning the attempt of a QCD-scale axion. In some limit one recovers the ordinary light axion pseudo-scalar,  but it is accompanied by a light scalar, 
excluded by  fifth-force constraints. 

\section{The model and its generalizations}\label{sec:model}
We begin our discussion reviewing and extending the results in \cite{Murayama:2026ioh}.

\subsection{General PQ charges}

We extend the SM introducing a complex scalar field with unit charge under  a global U(1)$_{\rm PQ}$ symmetry under which the light quarks $q\equiv (u\,, d\,, s)$ have charge $n_q\equiv (n_u\,,n_d\,,n_s)$.
The effective Lagrangian invariant under PQ symmetry (up to anomaly) at energies above QCD confinement and below electro-weak symmetry breaking reads 
\begin{equation}
\label{eq:LQCD}
\begin{split}
\mathscr{L}&=  -\frac{1}{4} G^a_{\mu\nu} G^{a, \mu\nu} + \theta \frac{g_s^2}{32\pi^2} G^a_{\mu\nu} \tilde G^{a, \mu\nu} + i
 \sum_{q=u,d,s} \bar{q}  \slashed{D} q \\
&+f_{\varphi}^2 \bigg[ |\partial_\mu \varphi|^2 - m^2_\varphi |\varphi|^2- \lambda f_\varphi^2 |\varphi|^4\bigg]-  \sum_{q=\{u,d,s\}} m_q
 (\sqrt{2}\varphi)^{n_q} \bar q_L q_R + \text{h.c.} 
 \end{split}
\end{equation}
For later convenience we introduced the dimension-less $\varphi$ complex field with decay constant $f_\varphi$. 
Field re-phasings allow to make $m_q$ real and positive.
If one $n_q=0$, the corresponding quark $q$ acquires the usual mass term $m_q$.
If instead $n_q\neq 0$ it acquires a coupling to $\varphi$.
The PQ symmetry gets spontaneously broken by the QCD chiral condensate after confinement:
quarks develop a condensate 
$ \langle \bar q_{Li} q_{Rj}\rangle=- \delta_{ij} B_0f_\pi^2/2$,
triggering a vacuum expectation value for $\varphi$.
The mass parameter $m_\varphi^2$ will be fixed such that $\langle\varphi\rangle  \equiv 1/\sqrt{2}$ in the vacuum. 
As a consequence, quarks acquire an effective mass $m_q$ from their coupling to $\varphi$.
We write it
\begin{equation}
m_q = \kappa_q \frac{f_\varphi}{\sqrt{2}}\,.
\label{eq:yuk}
\end{equation}
The model \cite{Murayama:2026ioh} corresponds to $n_u=1$ and $n_d=n_s=0$.  In this case $\kappa_u$ is a Yukawa coupling and the quark condensate induces a PQ-breaking tadpole. We consider a general PQ charge assignment of quarks.
PQ charges $|n_q|>1$ correspond to a non-renormalizable interaction
(with $\varphi\to\varphi^*$ if $n_q<0$).
The case $n_q=2$ produces a PQ-breaking negative contribution to %$\varphi$ 
the squared mass of the real/radial part of $\varphi$.

We assume that  the mass term $m_\varphi^2$ and the quartic $\lambda$ are positive, so that $\varphi$ acquires a VEV only after confinement and chiral symmetry breaking.  
%Despite that the Goldstone boson of PQ arises only below the QCD scale, the anomaly structure of U$(1)_{\rm PQ}$ is  determined. 
U$(1)_{\rm PQ}$ is anomalous under QCD when the charges satisfy $\sum_q n_q\neq 0$.  The axionic coefficient $E/N$ 
(in the standard notation, defined e.g. in~\cite{2003.01100})
and the domain wall number $N_{\rm DW}$ are 
\be
\frac{E}{N}\equiv \frac{2}{3} \frac{(4 n_u+n_d+n_s)}{\sum_q n_q
} \,,\quad\quad N_{\rm DW}=2N= \bigg|\sum_q n_q\bigg|\,.
\ee 
%On the minimum, the quarks charged under PQ  acquire an effective mass 
%so that the quark couplings in eq. (\ref{eq:LQCD}) can be written as,
%\be
%-\sum_{q=\{u,d,s\}} m_q (\sqrt{2}\varphi)^{n_q} \bar q_L q_R + \hbox{h.c.}
%\ee
%for both $n_q=0$ and $n_q\neq0$.

The extra scalars lie around the GeV scale,
so one may hope they exist hidden inside the many heavy QCD resonances. 
To compute their masses, it is convenient to
write the chiral Lagrangian including the external PQ field $\varphi$. In the massless-up quark solution the $\eta'$ acts as an axion so we include $\eta'$ in the chiral Lagrangian (we can neglect the difference $f_{\eta'}\approx 1.3 f_\pi \neq f_\pi$) writing down the leading term for the meson matrix $U=\exp(i\sqrt2\Pi/f_\pi)$  that transforms as $U\to L U R^\dag$ under U(3)$_L\otimes {\rm U}(3)_R$\footnote{We note that  the description in terms of the chiral Lagrangian is only valid at $E\lesssim$ GeV. Since in the relevant regime $\varphi$ will be heavier than GeV, it could be integrated out and matched to appropriate operators constructed with $U$. We perform this computation in section \ref{sec:EFT}. }
\begin{equation}\label{eq:chiral+phi}
\begin{split}
\mathscr{L}_{\pi\varphi}&=f_{\varphi}^2 \bigg[ |\partial_\mu \varphi|^2 - m^2_\varphi |\varphi|^2- \lambda f_\varphi^2 |\varphi|^4\bigg] \\
&+\frac{f_\pi^2}{4} \tr[D_\mu U D^\mu U^\dagger]+ \frac {f_\pi^2 B_0} {2}\tr [M_q U^\dagger + M_q^\dagger U]
- \frac{m_1^2 f_\pi^2}{12} \left[ \left(-\frac{i}{2} \ln \det U +\hbox{h.c.} \right) - \theta  \right]^2,
%\frac{m_1^2 f_\pi^2}{\MR{12}} [(\AT{-i \ln \det U -\theta})^2 +\hbox{h.c.}].
\end{split}
\end{equation}
where the last term reproduces the QCD anomaly to leading order in $1/N$. Through a PQ transformation $\theta$ can be set to zero explaining why the strong interactions are CP invariant in the low energy effective theory.

This Lagrangian captures the low-energy behaviour of the full theory of eq.~\eqref{eq:LQCD}.
The quark mass matrix 
\be
M_q =\mathrm{diag}\left[m_u (\sqrt{2}\varphi)^{n_u},\,m_d (\sqrt{2}\varphi)^{n_d},\,m_s (\sqrt{2}\varphi)^{n_s}\right]
\ee
depends on the dynamical field $\varphi$.
With this expression the mass matrix transforms as a spurion of U(3)$_L\otimes\,{\rm U}(3)_R$ as $M_q \to L M_q R^\dag$, which explains the appearance of the term $\tr[M_q U^\dag]$. We follow the conventions in~\cite{Pich:1991fq}, 
where $f_\pi\approx 92 \, \mathrm{MeV}$, $B_0=m_{\pi^\pm}^2/(m_u+m_d)$, 
$m_1^2$ is a proxy for the squared mass of $\eta'$ from the axial anomaly,
and the meson matrix $\Pi$ is 
\be\label{eq:PI}
\frac{\Pi}{f_\pi} = \begin{pmatrix}
\displaystyle \frac{\pi^0}{\sqrt{2}} + \frac{\eta}{\sqrt{6}} +\frac{\eta'}{\sqrt{3}}
& \pi^+ 
& K^+ \\[6pt]
\pi^- 
& \displaystyle -\frac{\pi^0}{\sqrt{2}} + \frac{\eta}{\sqrt{6}} +\frac{\eta'}{\sqrt{3}} 
& K^0 \\[6pt]
K^- 
& \bar K^0 
& \displaystyle -\frac{2\,\eta}{\sqrt{6}}+\frac{\eta'}{\sqrt{3}}
\end{pmatrix}
\ee
in terms of dimension-less fields
related to the canonically normalized fields by 
$\pi^0_\mathrm{can} = f_\pi \pi^0$, etc.

The mass spectrum of the model is then obtained by minimizing the potential terms of eq.~\eqref{eq:chiral+phi}. The novelty is the presence of the new complex scalar $\varphi$,
that can be expanded in CP-even, $\sigma$, and CP-odd, $a$, components as
\begin{equation}\label{eq:exponent}
\varphi= \frac{1 +\sigma} {\sqrt{2}} e^{i a   }.
\end{equation}
The radial mode $\sigma$ is a scalar while the phase $a$ is a pseudo-scalar. 
For $\theta=0$ (that can always be chosen through a PQ transformation)
the vacuum expectation values of CP-odd scalars vanish due to CP invariance.
%The PQ symmetry allows to remove the $\theta$ angle and take the vacuum expectation values of the neutral mesons to be zero. 
%The  minimum condition comes from requiring $\sigma=0$ and $a = 0$ on the vacuum. 
%This condition implies a relation between  low-energy QCD parameters and the new axion parameters. 
$m_\varphi$ (or alternatively $f_\varphi$) is determined by the condition  $\sigma=0$ in the minimum so that, 
\be\label{eq:condition-vev}
m_\varphi^2= \frac{B_0 f_\pi^2}{f_\varphi^2}\sum_q m_q n_q - \lambda f_\varphi^2 .
\ee
Under the assumption that  both $m_\varphi^2$ and $\lambda$ are positive,\footnote{If the squared mass term is negative, PQ  symmetry will be spontaneously broken independently of the QCD dynamics, and the model reduces  to a DFSZ-like axion model~\cite{Zhitnitsky:1980tq,Dine:1981rt}.}
there is an upper bound on $f_\varphi$. 
For example, focusing on the model $\{n_u,n_d,n_s\}=\{1,0,0\}$, we get 
 \be 
 \label{eq:fphibound}
f_\varphi \lesssim \text{GeV} \, \left( \frac{5.2~ 10^{-5}}{\lambda} \right)^{1/4}\,.
\ee
A large decay constant can only be achieved for a tiny quartic. 
If $f_\varphi$ is well within this upper bound, the minimization condition is independent of $\lambda$, and $m_\varphi^2$ is determined by the mass of the heaviest quark with PQ charge. In the case of \cite{Murayama:2026ioh}, we have\footnote{Note that $m_\varphi$ is $\sqrt{2}$ times smaller than in~\cite{Murayama:2026ioh}. This discrepancy originates from the normalization of kinetic terms in the chiral Lagrangian.}
\be
m_\varphi^2= \frac{m_{\pi^\pm}^2 f_\pi^2}{f_\varphi^2}\frac{m_u}{m_u+m_d} =  (2.3 \, \kappa_u\,   \mathrm{GeV})^2 .
\ee
Assuming $\kappa_u\sim 1$ implies $f_\varphi \approx 3 \, \mathrm{MeV}$, 
while the limit $\kappa_u ,\lambda\sim 0$ realizes a scenario with large $f_\varphi$,
leading to an invisible axion window discussed in section~\ref{sec:invax}.

The mass of the radial mode is easily found
\be
m_\sigma^2 = m_\varphi^2 + 3\lambda f_\varphi^2= \frac{B_0 f_\pi^2}{f_\varphi^2}\sum_q m_q n_q + 2\lambda f_\varphi^2\,.
\label{eq:masssigma}
\ee
%We next compute the masses of the CP-odd scalars, $a$ and the mesons $\pi^0, \eta, \eta'$.

\subsection{CP-odd scalar mass matrix}
The mass of the pseudo-scalar $a$ is a little more complicated  due to mixing with QCD mesons $\pi^0, \eta, \eta'$.
Neglecting interactions with the CP-even $\sigma$, which amounts to set $\sigma=0$, the Lagrangian restricted to self-interactions of the CP-odd scalars around the minimum of \eqref{eq:condition-vev} is
\begin{equation}
\begin{split}
\label{eq:Letatheta}
 \mathscr{L}&=\frac {f_\varphi^2} 2(\partial a)^2+\frac {f_\pi^2} 2 [(\partial \pi^0)^2+ (\partial \eta)^2+ (\partial \eta')^2 ]
 -\frac{m_1^2}{2}  f_\pi^2 (\eta'-\theta)^2 - B_0 f_\pi^{\,2}\Bigg[
m_s \cos\!\left(a\,n_s + \frac{2\eta - \sqrt{2}\,\eta'}{\sqrt{3}}\right) +\\
& + m_u \cos\!\left(a\,n_u - \frac{\eta}{\sqrt{3}} - \sqrt{\frac{2}{3}}\,\eta' - \pi^0\right)
+ m_d \cos\!\left(a\,n_d - \frac{\eta}{\sqrt{3}} - \sqrt{\frac{2}{3}}\,\eta' + \pi^0\right)  
\Bigg].
\end{split}
\end{equation}
It is useful to see explicitly how the 
strong CP problem is solved. Under a PQ transformation $\bar q_L q_R \to  \bar q_L q_R e^{i\alpha Q_{\rm PQ}}$, 
with $Q_{\rm PQ} = \text{diag}(n_u,n_d,n_s)$,
the pseudo-scalar fields transform as: 
$a \to a - \alpha$ and $U \to e^{-i\alpha Q_{\rm PQ}} U$. In particular
\be
\pi^0 \to \pi^0 - \frac{\alpha}{2} (n_u - n_d) , 
\qquad 
\eta \to \eta - \frac{\alpha}{2\sqrt{3}} (n_u + n_d - 2 n_s) ,
\qquad 
\eta' \to \eta' - \frac{\alpha}{\sqrt{6}} (n_u + n_d + n_s) .
\ee
The $\cos$ terms 
in eq.~\eqref{eq:Letatheta} are invariant under the PQ symmetry.
So the PQ shift of $\eta'$ can be used to reabsorb the $\theta$ term, 
as long as $\sum_q n_q \neq 0$. 
A quark charged under PQ 
($n_q \neq 0$) induces a mixing between the pseudo-scalar mesons and $a$,
so that the canonical axion field is a linear combination of 
the pseudo-scalar mesons
%$\eta'$ 
and $a$, 
weighted respectively by $f_\pi$ and $f_\varphi$.

The  $4\times 4$ symmetric mass matrix $\mathcal{M}$ in the basis $\{a, \pi^0, \eta, \eta'\}$ is 
\begin{equation}\label{eq:MPS}
\resizebox{0.9\textwidth}{!}{$
\mathcal{M}^2=B_0 \begin{pmatrix}
\dfrac{ f_\pi^{2}}{f_\varphi^{2}}\sum_q m_q n_q^2
& \dfrac{ f_\pi}{f_\varphi} (m_d n_d-m_u n_u )
& \dfrac{ -f_\pi}{\sqrt{3}\, f_\varphi} ( m_u n_u+m_d n_d - 2 m_s n_s )
& \dfrac{-\sqrt{2} f_\pi }{\sqrt{3}\, f_\varphi}\sum_q m_q n_q
\\[4pt]

-
%\dfrac{ f_\pi (-m_d n_d + m_u n_u)}{f_\varphi}
&  m_u + m_d
& \dfrac{(m_u-m_d )}{\sqrt{3}}
& \sqrt{\dfrac{2}{3}}\, (m_u-m_d )
\\[4pt]

-
%\dfrac{f_\pi (m_d n_d  + m_u n_u- 2 m_s n_s)}{\sqrt{3}\, f_\varphi}
& 
-
& \dfrac{1}{3}(m_d + m_u + 4 m_s)
& \dfrac{\sqrt{2} }{3}(m_d + m_u- 2 m_s )
\\[4pt]

%\dfrac{ f_\pi (m_u n_u+m_d n_d + m_s n_s )\sqrt{2}}{\sqrt{3}\, f_\varphi}
-
& 
-
%\sqrt{\dfrac{2}{3}}\, (m_u-m_d )
& 
-
%\dfrac{\sqrt{2} }{3}(m_u+m_d - 2 m_s )
&\dfrac{m_1^{2}}{B_0} + \dfrac{2}{3}  ( m_u+m_d + m_s ) 
\end{pmatrix}
$}.
\end{equation}
The mixing of $a$ with the other states depends on the charges $n_{u,d,s}$ and on  $f_\pi/f_\varphi$. 
If all quark charges are set to zero $a$ is massless.
For large $f_\varphi \gg {\rm GeV}$, $a$ becomes weakly coupled to QCD as in the usual invisible axion theories. 
In general, $f_\varphi$ is given by eq.~\eqref{eq:condition-vev}. 
For charges $n_q\leq 1$, the (1,1) entry of the matrix can be rewritten as 
\be
\mathcal{M}^2_{11}= m_\varphi^2 + \lambda f_\varphi^2 = m_\sigma^2  - 2\lambda f_\varphi^2.
\ee
In the limit of vanishing $\lambda$, $\mathcal{M}_{11}^2$ is equal to $m^2_\sigma$. 
Depending on the size of the correction induced by mixing with the other mesons, 
the axion-like scalar could be slightly heavier or lighter than the radial mode. 
For small $f_\varphi$ we expect it to be slightly heavier, while the opposite occurs for large $f_\varphi$.
As a check, the determinant of $\mathcal{M}^2$ is
\begin{equation}
\det\mathcal{M}^2=\frac{4 f_\pi^2 }{3 f_\varphi^2} B_0^3  m_u m_d m_s (n_u+n_d + n_s )^2 m_{1}^2.
\end{equation}
It vanishes when one of the quark masses is zero, 
as well as if PQ is non-anomalous ($\sum_q n_q=0$) leading to a massless axion. 
Before analyzing the mass spectrum in various models in section \ref{sec:QCDscaleAxion},
we consider a possibly relevant higher-order correction in the chiral Lagrangian.

\subsection{Kaplan-Manohar ambiguity of the chiral Lagrangian}
Going to higher orders in the chiral expansion, extra operators contribute to the chiral Lagrangian,
and can be relevant for our discussion.
Of particular relevance is 
\begin{equation}\label{eq:KMO}
\Delta \mathscr{L}=\frac{B_0 f_\pi^2}{2}x_{\rm KM} \mathcal{O}_{\rm KM},\qquad
  \mathcal{O}_{\rm KM} \equiv
-\frac12\left[ \tr (M_q U^\dagger M_q U^\dagger) -
    \tr (M_q U^\dagger)^2 + \hbox{h.c.}\right].
   % = {\cal O}_8-\frac{{\cal O}_6+{\cal O}_7}{2}
\end{equation}
Such NLO term is  equivalent to shifting the quark mass matrix as
\begin{equation}
M_q\to   M_q|_{\rm KM}  =M_q  + x_{\rm KM} (M_q^{-1})^\dag\det (M_q^\dag) \det U 
\end{equation}
in the LO operator $\tr(M_q U^\dag)+\hbox{h.c.}$~\cite{Kaplan:1986ru},
in view of the Cayley-Hamilton identity $2\det M_q \,\tr M_q^{-1}=(\tr M_q)^2-\tr M_q^2$. The new term has the same symmetry transformations as $M_q$.
The presence of $\det U =e^{i\sqrt{6} \eta'}$ guarantees that the two terms have the same transformation also under the axial U(1)$_{\rm A}$.
The constant $x_{\rm KM}$ is approximately determined by lattice computations.
This is known as the Kaplan-Manohar (KM) ambiguity~\cite{Kaplan:1986ru}, since it complicates the extraction of bare quark masses from meson data. In particular, it replaces $m_u \to m_u + x_{\rm KM} m_d m_s$.
This was considered as a loophole for reviving the massless up-quark solution to the strong CP problem, before lattice calculations established that $m_u\neq 0$. 
%\cite{Durr:2015dna}. 
%Since the up is massless up to scales of order $\Lambda_{\rm QCD}$, we take the point of view to include this in the chiral Lagrangian by means of the modification to the term in $B_0$ as
%\be
%- \frac{f_\pi^2B_0}{2} \tr[M_q U^\dag + x_{\rm KM} (M_q^{-1})^\dag\det (M_q^\dag) (\det U) U^\dag]  + \text{h.c.} , 
%\ee
In our case the KM `ambiguity' effectively redefines the quark mass matrix into
\be\label{eq:M_KM}
M_q|_{\rm KM}=\mathrm{diag} \left(\begin{array}{c}
m_u (\sqrt{2}\varphi)^{n_u} + x_{\rm KM}m_s m_d (\sqrt{2}\varphi^*)^{n_d+n_s}e^{i\sqrt{6}\eta'} 
\\  m_d (\sqrt{2}\varphi)^{n_d} + x_{\rm KM} m_u m_s (\sqrt{2}\varphi^*)^{n_u+n_s}e^{i\sqrt{6}\eta'} 
\\  m_s (\sqrt{2}\varphi)^{n_s} + x_{\rm KM} m_u m_d (\sqrt{2}\varphi^*)^{n_u+n_d}e^{i\sqrt{6}\eta'} \end{array}
\right)\,.
\ee
The $\eta'$ phase  is necessary to reproduce the transformation under U(1)$_{\rm PQ}$ and corrects the meson mass matrix introducing an extra mixing with $\eta'$.
In addition to shifting the quark masses, the KM term modifies the
dependence on $\sigma$ when $n_u\neq n_d$.
Its implications will be discussed in the following.

\section{QCD-scale axion}\label{sec:QCDscaleAxion}
The chiral Lagrangian at LO predicts that pions form a degenerate triplet, with the observed mass difference arising primarily from electromagnetic effects~\cite{1604.07112,2108.05311,2202.11970}
\begin{equation} \label{eq:piem}
\Delta_{\rm em} (m_{\pi^0}^2 - m_{\pi^\pm}^2)\approx -0.0013\, {\rm GeV}^2 \sim -3\frac{\alpha_{\rm em} \Lambda^2}{4\pi},  
\end{equation}
where $\Lambda\sim m_\rho$.  Adding a QCD-scale axion generically breaks the isospin symmetry, ruining this successful prediction. 
This can be seen integrating out $\varphi$ at tree level. In  appendix \ref{sec:loop} we present a related computation of loop corrections induced by $\varphi$ analogous to the electromagnetic splitting.

\subsection{Tree-level corrections to meson masses}
This class of models predicts important modifications to  meson masses.
To simplify formulas, we now focus on $n_s=0$.
In the limit where the $a$ is the heaviest state, 
i.e.~$m_\varphi>$ GeV, we can integrate it out. 
This adds to the QCD squared mass matrix of neutral mesons the following additive correction
\begin{equation}\label{eq:DeltaMM}
\Delta M^2=B_0
\begin{pmatrix}
-\dfrac{(m_d n_d - m_u n_u)^2}{m_d n_d^2 + m_u n_u^2}
&
\dfrac{m_d^2 n_d^2 - m_u^2 n_u^2}
{\sqrt{3}\,(m_d n_d^2 + m_u n_u^2)}
&
\dfrac{\sqrt{\frac{2}{3}}\,(m_d^2 n_d^2 - m_u^2 n_u^2)}
{m_d n_d^2 + m_u n_u^2}
\\[1.2em]

\dfrac{m_d^2 n_d^2 - m_u^2 n_u^2}
{\sqrt{3}\,(m_d n_d^2 + m_u n_u^2)}
&
-\dfrac{(m_d n_d + m_u n_u)^2}
{3\,(m_d n_d^2 + m_u n_u^2)}
&
-\dfrac{\sqrt{2}\,(m_d n_d + m_u n_u)^2}
{3\,(m_d n_d^2 + m_u n_u^2)}
\\[1.2em]

\dfrac{\sqrt{\frac{2}{3}}\,(m_d^2 n_d^2 - m_u^2 n_u^2)}
{m_d n_d^2 + m_u n_u^2}
&
-\dfrac{\sqrt{2}\,(m_d n_d + m_u n_u)^2}
{3\,(m_d n_d^2 + m_u n_u^2)}
&
-\dfrac{2(m_d n_d + m_u n_u)^2}
{3\,(m_d n_d^2 + m_u n_u^2)}
\end{pmatrix}.
\end{equation}
The spectrum of charged  mesons is not modified at tree level,
since the new state is neutral and there is no mixing with new states. 
Using $m_s\gg m_{u,d}$, the 11 entry of the mass matrix leads to the pion mass splitting
\begin{equation}
\label{eq:Deltampi0pip}
\frac{\Delta (m_{\pi^0}-m_{\pi^\pm})}{m_\pi} \approx -\frac12\dfrac{(m_u n_u-m_d n_d )^2}{(m_d n_d^{2}  + m_u n_u^{2})(m_u+m_d)}\,, 
\end{equation}
Other mesons are affected too.
In particular, the $\eta$ is lighter than in the SM.

\subsection{Model with axion coupled to the up quark}
Murayama~\cite{Murayama:2026ioh} considered the case where  the up quark only is PQ-charged,
$n_u=1$ and $n_d=0$.
In this case one finds (neglecting electromagnetic corrections)
\begin{equation}
m_\varphi \approx 2.3 \kappa_u{\rm GeV},\qquad
%\frac{\Delta m_{\pi^0}}{m_{\pi}}\approx \frac {m_u}{2(m_u+m_d)}\approx -0.15
m_{\pi^0}^2\approx m_d B_0,
\qquad
m_{\pi^\pm}^2 \approx B_0 (m_u+m_d), 
\end{equation}
that is equivalent to eq.~(\ref{eq:splitting1}).
A GeV-scale $\varphi$ needs a relatively large Yukawa coupling, $\kappa_u \gtrsim 0.5$, not much below the perturbativity limit.
%On the other hand $\kappa_u$ should be smaller than the QCD couplings. This essentially predicts,
%\begin{equation}
%\kappa \approx 0.5
%\end{equation}
The KM ambiguity helps to alleviate the problem with pion masses~\cite{Murayama:2026ioh},
but only mildly. It shifts the quark mass matrix into
\be\label{eq:M_KMu}
M_q|_{\rm KM}=\mathrm{diag} \left(\begin{array}{c}
m_u \sqrt{2}\varphi +  x_{\rm KM}  m_s m_d e^{i\sqrt{6}\eta'} 
\\  m_d  +  x_{\rm KM} m_u m_s \sqrt{2}\varphi^*e^{i\sqrt{6}\eta'} 
\\  m_s  + x_{\rm KM} m_u m_d \sqrt{2}\varphi^*e^{i\sqrt{6}\eta'} \end{array}
\right)\,.
\ee
The KM contribution does not modify $\langle{\varphi}\rangle$. 
The problematic correction to the pion mass then gets suppressed if the up entry of $M_q|_{\rm KM}$ is dominated by the KM shift, as it is $\varphi$-independent.  Assuming that the up quark mass induced by $\langle\varphi\rangle$
is a factor $R$ smaller than the measured up quark mass, we obtain
\begin{equation}
\frac{\Delta m_{\pi^0}}{m_{\pi^0}}\approx -\frac{0.15}R,
\qquad
m_\varphi \approx 2.3 \, \kappa_u \, \sqrt{R}\, {\rm GeV}.
\label{eq:splitting}
\end{equation}
Avoiding the pion mass problem requires $R\gtrsim 10$.
This corresponds to a value of $x_{\rm KM}$ much larger than
the value suggested by lattice computations, that favor a mild KM shift
and exclude $m_u=0$, corresponding to $R=\infty$~\cite{hep-lat/0212009,2411.04268,PDG}. Thus it seems unlikely that the KM ambiguity solves the pion mass problem.

If the model were viable, the new GeV-scale scalars should be identified
with some observed hadron resonances. Table~\ref{tab:eta_f_resonances} lists some candidates.
Decay rates of $a$ and $\sigma$ can  be estimated either from decays to quarks or to pions. 
Both computations are approximate and lead to similar estimates:
\begin{equation}
\Gamma(a\to 3 \pi)\approx 0.1 \kappa^2_u m_a  ,\qquad
\Gamma(\sigma \to \pi\pi)\approx  \frac {\kappa^2_u}{8\pi} m_\sigma .
\end{equation}
The relatively narrow $a$ could be identified with the observed $\eta(1295)$,
as its mass is $m_a \sim 1.3 \,{\rm GeV}$ and its width is $\Gamma_a\sim 25\,{\rm  MeV}$
for $\kappa_u\sim 1/2$ 
(we fixed factors of 2 compared to \cite{Murayama:2026ioh}).
The broad $\sigma$ can be more easily hidden inside CP-even QCD resonances. 
In case the pion mass problem were eventually solved, a dedicated analysis of $a$ and $\sigma$ decays, similar to the one performed for GeV-scale axion-like particles in e.g.~\cite{1811.03474,2406.11948,2501.04525,2505.24822,2506.15637}, would be needed.

\begin{table}[t]
\centering
\renewcommand{\arraystretch}{1.2}
\begin{tabular}{lccc}
State & $J^{PC}$ & Mass [MeV] & Width [MeV] \\
\hline
$\eta(547)$      & $0^{-+}$ & $547.9 \pm 0.0$   & $(1.31 \pm 0.05)\times 10^{-3}$ \\
$\eta'(958)$     & $0^{-+}$ & $957.8 \pm 0.1$   & $0.198 \pm 0.009$ \\
$\eta(1295)$     & $0^{-+}$ & $1294 \pm 4$      & $55 \pm 5$ \\
$\eta(1405)$     & $0^{-+}$ & $1409 \pm 2$      & $51 \pm 7$ \\
$\eta(1475)$     & $0^{-+}$ & $1476 \pm 4$      & $85 \pm 9$ \\
$\eta(1760)$     & $0^{-+}$ & $1756 \pm 9$      & $96 \pm 70$ \\
$\eta(2225)$     & $0^{-+}$ & $2221 \pm 8$      & $185 \pm 40$ \\
\hline
$f_0(500)$       & $0^{++}$ & $400$--$550$      & $400$--$700$ \\
$f_0(980)$       & $0^{++}$ & $990 \pm 20$      & $40$--$100$ \\
$f_0(1370)$      & $0^{++}$ & $1200$--$1500$    & $200$--$500$ \\
$f_0(1500)$      & $0^{++}$ & $1505 \pm 6$      & $109 \pm 7$ \\
$f_0(1710)$      & $0^{++}$ & $1723 \pm 6$      & $135 \pm 8$ \\
$f_0(2020)$      & $0^{++}$ & $1992 \pm 16$     & $442 \pm 60$ \\
$f_0(2200)$      & $0^{++}$ & $2189 \pm 13$     & $238 \pm 50$ \\
$f_0(2330)$      & $0^{++}$ & $2325 \pm 35$     & $200 \pm 70$ \\
\end{tabular}
\caption{\it Isoscalar pseudoscalar ($\eta$) and scalar ($f_0$) resonances in QCD with their quantum numbers, masses, and total decay widths. Broad states have large systematic uncertainties.}
\label{tab:eta_f_resonances}
\end{table}

\subsection{Model with axion coupled to up and down quarks}\label{sec:ud}
Given that the pion mass splitting originates from the breaking of the SU(2) isospin symmetry, 
it is tempting to assume that both up and down quarks are PQ charged.
For negligible $\lambda$ the mass of $\sigma$ becomes
\begin{equation}
m_\varphi =1.6\,\kappa_d \, {\rm GeV}
 \sqrt{n_u\frac{m_u}{m_d}+n_d}\,.
\end{equation}
The pion mass splitting decreases to 7$\%$ for $n_u=n_d=1$. The reason for this is that the breaking of isospin is milder given that $m_d\sim 2 m_u$.
This is still excluded.
Moreover in this case the KM ambiguity is of little help, as it modifies 
in a similar way the two entries of the quark mass matrix
\be\label{eq:M_KMud}
M_q|_{\rm KM}=\mathrm{diag} \left(\begin{array}{c}
m_u \sqrt{2}\varphi +  x_{\rm KM} m_s m_d \sqrt{2}\varphi^*e^{i\sqrt{6}\eta'} 
\\  m_d \sqrt{2}\varphi +  x_{\rm KM}  m_u m_s \sqrt{2}\varphi^* e^{i\sqrt{6}\eta'} 
\\  m_s  + x_{\rm KM}  m_u m_d (\sqrt{2}\varphi^*)^2 e^{i\sqrt{6}\eta'} \end{array}
\right)\,.
\ee
A  cancellation of the pion mass shift would take place for $n_u=2$ and $n_d=1$ (see~eq.~\eqref{eq:Deltampi0pip}), as data allow for $m_d/m_u=2$ within uncertainties. 
However $n_u=2$ means that the up-quark mass arises
from a dimension-5 operator that gets strongly coupled at
energies just above $m_u$, well below the QCD scale. 
This would require a UV completion at that scale,
with extra particles that would be problematically light.

\subsection{Model with axion coupled to the strange quark}
We briefly comment on the possibility that only the strange quark has a PQ charge. 
This would have the merit of preserving the isospin $\mathrm{SU}(2)$ symmetry of the $u,d$ sector. However one predicts
\begin{equation}
\label{eq:mphistrange}
m_\varphi = 
 \left( \frac{B_0 f_\pi^2}{2m_s} \right)^{1/2} \kappa_s
\approx 0.36~\text{GeV}\,\kappa_s
\end{equation}
for the radial mode. Upon diagonalization, the lightest pseudo-scalar turns out to be even lighter (this can be understood as the mass of the strange quark does not enter the mass matrix of neutral mesons as discussed in the next section), $m_a \approx 0.038~\text{GeV}$ for $\kappa_s=1$. 
Such a light axion is excluded by old collider data, and by flavour constraints, most notably by $D\to \pi a$, generated at one loop via $W$ exchange and sourced by the $a\,\bar s\, i\gamma_5 s$ coupling. Moreover, the pion-mass issue is effectively shifted to the $\eta$ sector: the $\eta$ mass receives a sizeable correction,
$\Delta m_\eta/m_\eta \approx +15\%$ for $\kappa_s=1$,
which is difficult to reconcile with data. 

Extra tentative models {with multiple PQ scalars}
are discussed in appendix~\ref{sec:appfail}. 
One might even consider the possibility of adding a PQ-neutral scalar just to make the $\pi^0$ heavier. In section~\ref{sec:EFT} we discuss the general reason that prevents realizing a viable model.

\subsection{Loop corrections to pion masses}
All results receive corrections that become large if $\kappa$ is large. 
Here we attempt an estimate considering radiative corrections induced by the new coupling not included in lattice computations. One-loop corrections to pion masses directly affect the dominant diagonal entries of the meson mass matrix in eq.\eq{MPS}.
As a consequence there could be a controllable regime where the one-loop corrections cancel the too large tree-level correction to $m_{\pi^0}-m_{\pi^\pm}$,  arising from off-diagonal entries in the meson mass matrix.
In appendix \ref{sec:loop} we discuss  how the pion splitting could be precisely computed using QCD data in the regime where $\kappa$ is perturbative.
We here heuristically derive a similar result 
by computing the renormalization of the meson masses in the chiral Lagrangian effective theory, including $\varphi$.
In addition to isospin-symmetric vertices, it contains a $\sigma \pi^0 \eta^{(\prime)}$ vertex that only affects the $\pi^0$ mass.
The linear coupling of $\varphi$ to QCD mesons can be written in terms of $\varphi$-dependent meson mass matrix, obtaining the 1-loop contribution as
\begin{equation}
\Delta M^2_{ij}=- (\partial_\varphi M^2_{ik})\left[ \int \frac{d^4Q}{(2\pi)^4} \frac 1 {Q^2+m_\varphi^2} \frac 1 {Q^2+m^2}\right]_{kl} (\partial_\varphi M^2_{lj}),
\end{equation}
where latin indices label QCD mesons.
The integrand is positive and the integral is logarithmically UV divergent. 
More importantly, we cannot apply the chiral Lagrangian at large $Q\gtrsim$ GeV so we estimate the contribution with a cut-off at $\Lambda\sim {\rm GeV} $. Assuming $m_\varphi\gtrsim \Lambda$  and
neglecting for simplicity off-diagonal terms in the mass matrix, we find
\begin{equation}
\Delta M^2_{ij}\approx -\frac {(\partial_\varphi M^2_{ik})} {16\pi^2 m_\varphi^2}\left[\Lambda^2-  M^2_{kk} \ln \frac {\Lambda^2+ M_{kk}^2}{M_{kk}^2} \right](\partial_\varphi  M^2_{kj}).
\end{equation}
To leading order for $n_u=1$ we find,
\begin{equation}
\Delta M^2|_{\rm neutral}\approx -\frac{\kappa^{2}B_0^2\Lambda^2}{(4 \pi m_{\varphi})^2}
\begin{pmatrix}
2 & \frac 2 {\sqrt{3}} & 2\sqrt{\frac 2 3} \\
2 {\sqrt{3}} & \frac 2 3 & \frac {2\sqrt{2}}3\\
2\sqrt{\frac 2 3} & \frac {2\sqrt{2}}3 & \frac 4 3
\end{pmatrix},\qquad
\Delta M^2|_{\rm charged}\approx -\frac{\kappa^{2}B_0^2\Lambda^2}{(4 \pi m_{\varphi})^2}\begin{pmatrix}
1 & 0 & 0 \\
0 & 1 & 0 \\
0 & 0 & 0
\end{pmatrix}.
\label{eq:estloop}
\end{equation}
This splitting, when extrapolated to $m_\varphi>\Lambda$ agrees, including the sign, with the result of the tree level matching eq.~\eqref{eq:DeltaMM} 
upon identifying $\Lambda= 4\pi f_\pi$. This however should not be considered as a new contribution to the splitting but rather as an estimate of the uncertainty. In the appendix we present a more precise computation that could in principle allow to determine the correction in terms of QCD correlators.

\section{Heavy axion effective chiral theory}\label{sec:EFT}
In this section we focus on the regime where the new PQ scalars ($a$ and $\sigma$ in the minimal model) 
are heavy enough
that they could be hidden among QCD resonances with GeV mass.
This regime needs $n_s=0$ and $n_{u,d}\le1$ in the language of the previous section.
We here exploit the (mild) separation of scales  
\begin{equation}
    m_{\sigma, a} \mathrel{\raisebox{-1ex}{$\overset{\gg}{\sim}$}} m_{\pi,K,\eta}
\end{equation}
to integrate out the PQ scalars,
obtaining a modified chiral Lagrangian for the light mesons.
Namely we capture the effect of the heavy field 
by matching the full theory of eq.~\eqref{eq:LQCD} at $m_\varphi$ 
to a new chiral Lagrangian augmented by operators --- invariant under PQ --- generated by the new scalar dynamics. 
This allows to exploit the symmetries to constrain the form of extra effects, including loop effects.
The low energy degrees of freedom are again the $\Pi$ fields of eq.~\eqref{eq:PI}. The $\eta'$ is again kept in the effective field theory (EFT) because of its special role for the PQ symmetry.
The meson interactions  differ from the usual chiral Lagrangian, and even include different isospin-breaking terms. 
Indeed, at the matching scale, formal invariants under the full symmetry 
now involve two spurions: 
\begin{itemize}
 \item   The usual mass matrix of  quarks, with masses now present only for PQ-neutral quarks:
 \begin{equation}
 \tilde M_q=\mathrm{diag}(m_u (1-n_u), m_d (1-n_d), m_s).
 \end{equation}
 
 \item The axion couplings to quarks, proportional to the spurion
\begin{equation}
I_{\rm PQ}=\mathrm{diag}(\kappa_u n_u, \kappa_d n_d, 0).
\end{equation}
\end{itemize}
It has the same transformation properties $I_{\rm PQ} \to L I_{\rm PQ} R^\dag$
of $\tilde M_q$  under  U(3)$_L\otimes\,{\rm U}(3)_R$,
but it transforms as 
\begin{equation}
I_{\rm PQ} \to I_{\rm PQ} e^{-i\alpha Q_{\rm PQ}},\qquad Q_{\rm PQ}={\rm diag}(n_u, n_d,0) 
\end{equation}
under the PQ symmetry that acts on quarks as
$\bar q_L q_R \to e^{i\alpha Q_{\rm PQ}} \bar q_L q_R$,
and on mesons as $U \to U e^{-i\alpha Q_{\rm PQ}}$.
Integrating $\varphi$ out at the level of the QCD action generates
PQ-invariant $|\bar q_L q_R|^2$ operators,
while $n_q >1$ would lead to higher quark powers.
The modified chiral EFT is constrained by
the invariance under the PQ symmetry, and has the form
\be\label{eq:SMEFT}
\mathscr{L}_{\rm eff} = \mathscr{L}_{\rm QCD}(\tilde M_q) + \sum_i C_i \mathcal{O}_i^{\rm PQ} .
\ee
The first term, $\mathscr{L}_{\rm QCD}(\tilde M_q)$, is the chiral Lagrangian 
with quark masses given by $\tilde M_q$. 
As in eq.\eq{chiral+phi}, it includes the term
$\ {f_\pi^2 B_0} \tr [\tilde{M}_q U^\dagger + \tilde M_q^\dagger U]/2$
at leading $p^2$ order in the chiral expansion, and terms such as  $\mathcal{O}_{\rm KM}$ at higher order.
The second is a sum of operators with at most two derivatives  generated at the scale $m_\varphi$. Focusing on $n_q=0,1$, 
\begin{itemize}
\item One operator appears at $p^4$ order in the chiral expansion
\be
\label{eq:OPQ1}
\mathcal{O}_1^{\rm PQ} = \tr[I_{\rm PQ}^\dag U]\tr[I_{\rm PQ} U^\dag].
%c\,,\qquad    \mathcal{O}_2^{\rm PQ} = \mathcal{O}^{\rm KM}|_{\tilde M_q}
\ee

\item Two extra operators contribute at next-to-leading $p^6$ order,
\be\label{eq:basis}
\mathcal{O}_2^{\rm PQ} = \tr[I_{\rm PQ}^\dag\partial_\mu U]
\tr[I_{\rm PQ}\partial^\mu U^\dag], 
\qquad
\mathcal{O}_3^{\rm PQ} = \tr[I_{\rm PQ}^\dag (\partial_\mu U) ( \partial^\mu U)^\dag I_{\rm PQ}].
\ee
\end{itemize}
In this language, the strong CP problem is solved because
the operators  are invariant under a global PQ transformation of $U$, 
that arises when $\tilde M_q$ has at least a zero eigenvalue. 
Therefore the total effective action is PQ invariant up to the anomaly term.

\medskip

The  model \cite{Murayama:2026ioh} is now encoded in the coefficients $C_i$. 
Integrating out the PQ scalar at tree level gives  
\be\label{eq:coefficients}
C_1= \frac14 \frac{f_\pi^4 B_0^2}{m_\varphi^2},
\qquad
%C_2 = +\frac{B_0f_\pi^2}{2}x_{\rm KM}\,,\qquad
C_2 = \frac14\frac{f_\pi^4 B_0^2}{m_\varphi^4},
\qquad 
C_3=0.
\ee
At loop level one expects sizable corrections in light of the large coupling $\kappa$.
In the standard power counting of chiral perturbation theory, the coefficient $C_1$ would be suppressed by $m_\pi^2/(4\pi f_\pi)^2$, giving a small contribution to pion masses.
However the Murayama theory predicts a larger $C_1$ because ${\cal O}_1^{\rm PQ}$ reproduces the effect of up quark mass. So the theory breaks the  usual chiral expansion in powers of $p^2$. Unlike the leading order $p^2$ operator, 
$p^4$ operators (including ${\cal O}_1^{\rm PQ}$) contribute to pion masses in a way that breaks SU(2) isospin. 

\subsection{Pion masses}
This effective theory allows us to compute the meson masses,
and see why they differ from the QCD values.
%It is then illuminating to compute the mass matrix of the CP-odd neutral mesons $(\pi^0,\eta,\eta')$, as $C_1$ and $C_2$ will make their appearance. 
%We focus on the neutral sector, but we also retain information about the mass of $\pi^\pm$. 
The derivative operators ${\cal O}_{2,3}^{\rm PQ}$ 
act as a wave-function renormalization of the kinetic terms of the pions, and as such they affect the mass eigenvalues (after canonical normalization including rotation and rescaling)
giving sub-leading corrections in $1/m_\varphi^2$.
%This calculation will allow us to fix the parameters of \eqref{eq:SMEFT}, the ones in $\tilde M_q$ and $I_{\rm PQ}$ and $C_{i}$.
Using $m_s\gg m_{u,d}$, the masses of the neutral and charged pions are
\begin{eqnarray}\label{eq:masses-EFT}
m_{\pi^0}^2&\approx & B_0 \tilde{m} 
\bigg(1 + \frac{x_{\rm KM} m_s} {2}\bigg)+
%\frac{C_{\rm KM} m_s \tilde{m}}{f_\pi^2}+
\frac{8 C_1 \kappa_u \kappa_d n_u n_d}{f_\pi^2} ,\\
\label{eq:masses-EFT2}
m_{\pi_\pm}^2&\approx& B_0 \tilde{m} \bigg(1 + \frac{x_{\rm KM} m_s} {2}\bigg) 
%\frac{C_{\rm KM} m_s \tilde{m} }{f_\pi^2} 
+\frac{2C_1(\kappa_u n_u +\kappa_d n_d)^2}{f_\pi^2}, 
%\kappa_d n_d)^2+C_3(\kappa_u n_u - \kappa_d n_d)^2]}{f_\pi^2}.
\end{eqnarray}
where $\tilde{m}\equiv m_u(1-n_u) + m_d (1-n_d)$. 
We see that $C_1$ breaks isospin at the level of the mass spectrum.
For the charged pion one finds the standard result $m_{\pi^\pm}^2=B_0(m_u+m_d)$ (plus KM corrections) for $C_1$ as in eq.\eq{coefficients}. The  pion mass difference is
\be
m_{\pi^0}^2-m_{\pi_\pm}^2= -2\frac{C_1}{f_\pi^2}(\kappa_u n_u - \kappa_d n_d)^2
\ee
in agreement with our previous results in eq.~\eqref{eq:Deltampi0pip},
where here $n_{u,d}\le1$.
The EFT view clarifies in a more general way why it is difficult to keep the pions degenerate. Pion masses change in a qualitative way compared to QCD because they arise from a different operator, that breaks isospin invariance.
The correction to pion masses is numerically comparable to the effect of setting $m_u=0$ in QCD.

A few comments are in order. The KM ambiguity can help only partially, as (despite some similarity)
${\cal O}_1^{\rm PQ}$ differs from the KM operator in eq.\eq{KMO}. 
Isospin SU(2) could perhaps be recovered by adding a SU(2) multiplet of extra scalars at the QCD scale, but the electrically charged components would be excluded.
Loop corrections can become more significant, since $C_1$ violates the chiral expansion. Nevertheless, loop corrections cannot cancel the 
tree-level effect, as long as one operator is dominant.

\smallskip

As long as one operator dominates, its coefficient cannot be tuned to zero
in an extended theory with more GeV-scale PQ scalars.
One could consider an extension of the theory where
extra operators are added just to tune the pion masses to their observed  values. The EFT formalism shows however that this leaves other problems, in pion interactions.

\subsection{Pion interactions}
In QCD, the meson mass operator is invariant under SU(2) isospin.
Indeed, setting to zero the heavier mesons, its explicit form is
\begin{equation}\label{eq:mpiQCD}
\tr\,[M_q U^\dag + \hbox{h.c.}]= m_s + (m_u+m_d)\cos\pi
\qquad\hbox{where}\qquad
\pi^2 \equiv \pi^a \pi^a = (\pi^0)^2+ 2 \pi^-\pi^+ .
\end{equation}
This operator predicts a relation between the quadratic and quartic terms in the potential $V(\pi)$, namely between pion masses and couplings.
As a result the 4-pion scattering amplitude at tree level in QCD has
the Weinberg form
 \begin{equation}\label{eq:ApiQCD}
     \mathscr{A}(\pi_a\pi_b \to  \pi_c\pi_d)=
     \frac{1}{f_\pi^2}\bigg[
 \delta_{ab}\delta_{cd}  (s-m_\pi^2)+
 \delta_{ac}\delta_{bd}  (t-m_\pi^2)+
 \delta_{ad}\delta_{bc}  (u-m_\pi^2)  \bigg],
 \end{equation} 
that  vanishes in the soft pion limits, a feature known as 
Adler's zero. The $s,t,u$ terms arise from the pion kinetic term, while the constant terms arise from the pion mass term.
This term changes  if the higher-order chiral operator contributes to pion masses. Indeed, in the same limit, ${\cal O}_1^{\rm PQ}$ gives a different isospin-breaking
function of pions:
 \begin{equation}
      {\cal O}_1^{\rm PQ} =
 (n_u\kappa_u+n_d \kappa_d)^2\cos^2\pi +
 (n_u\kappa_u-n_d \kappa_d)^2\frac{(\pi^0)^2}{\pi^2} \sin^2\pi.
 \end{equation}
For some parameter values it leads to an extra local minimum, 
giving the cosmological signatures discussed in~\cite{1902.05933}.
Tuning $n_u \kappa_u-n_d \kappa_d$ to 0 allows to restore SU(2) invariance,
but with a different function of $\pi$ than in eq.\eq{mpiQCD}.
The quartic pion interactions, in addition to the QCD ones, are
 \begin{equation}
\mathscr{L}_{4\pi}=
\mathscr{L}_{4\pi}^{\rm QCD}
+ B_0 f_{\pi }^2\pi ^2  
\frac{6 \pi ^- \pi ^+ \left(m_d n_d+m_u n_u\right)^2 -(\pi ^0)^2 (m_d^2 n_d^2+m_u^2 n_u^2-14 m_d n_d m_u n_u)}{24 \left(m_d
   n_d+m_u n_u\right)}.
\end{equation}
% The quartic term in ${\cal O}_1^{\rm PQ}$ 
%  \begin{equation}
% \frac{4 }{3f_\varphi^2}\pi^2 \bigg[2 m_u m_d n_u n_d (\pi^0)^2+
% (m_u n_u +m_d n_d)^2 \pi^-\pi^+\bigg]
%  \end{equation}
Such terms contribute to pion scattering amplitudes, changing the
constant term in pion scattering amplitudes from the QCD value of eq.\eq{ApiQCD}.
The Murayama model with $n_d=0$ gives a vanishing correction to scatterings involving $\pi^0$
 only, but affects processes involving charged pions.
 For example, the simpler amplitudes that involve one pion mass only change into 
\begin{equation}
    \mathscr{A}(\pi^0\pi^0\to\pi^0\pi^0) =(4-3X_0) \frac{m_{\pi^0}^2}{f_\pi^2},\qquad
    X_0=\left\{\begin{array}{ll}
        1 & \hbox{if one $n_q=0$}\\
        4 & \hbox{if $n_u=n_d=1$}
    \end{array}\right. ,
\end{equation}
\begin{equation}
    \mathscr{A}(\pi^+\pi^-\to\pi^+\pi^-) = \frac{s+t- 2 X_\pm m_{\pi^\pm}^2}{f_\pi^2},\qquad X_\pm = 1 + 3 \frac{m_u n_u + m_d n_d}{m_u+m_d}.
\end{equation}
Here $X_{0,\pm}$ are the relevant $V''''/V''$ terms,
normalized such that the standard QCD result corresponds to $X_{0,\pm}=1$.
The difference in the pion cross section is of order unity in the non-relativistic limit
$s\to 4m_\pi^2$, $t\to 0$.
Such cross sections are measured with
few $\%$ precision to agree with the QCD values, see e.g.~\cite{1211.3026}.

% Pions interactions are also modified. In particular one generates,
% \begin{equation}
% \mathscr{L}_{\rm QCD}+
% \frac{1}{24}\, B_0\, f_{\pi}^{2}\, m_u \;
% \Big[
% -12\big(\pi^+\pi^-\big)^{2}
% -4\pi^+ \pi^-\pi_{0}^{2}
% +\pi_{0}^{4}
% \Big]\;
% \end{equation}
% New mixings and interactions with $\eta$ are also produced,
% \begin{equation}
% B_0\eta\frac{ f_{\pi}^{2}\, m_u\, \, \pi_0\,
% \big(-6 + (2\pi^+\pi^- +\pi_{0}^{2})\big)}
% {6\sqrt{3}}
% \end{equation}

\section{Light invisible axion}\label{sec:invax}
A heavy axion resembling a QCD meson would be a qualitatively new solution to the strong CP problem, but we could not cure its phenomenological problems.
We conclude by considering the same theory in the 
limiting regime where the axion is much lighter than the QCD scale,
so that the EFT of section~\ref{sec:EFT} is no longer applicable.
As $f_\varphi$ becomes large, the pseudo-scalar component $a$ (and its scalar companion $\sigma$) become lighter, producing an invisible axion scenario with the special property that QCD itself breaks the PQ symmetry, 
leading to a non-standard cosmological history. 
In this region of the parameter space the masses of the 
two scalars read (in the following $n_q \leq 1$)
\be
\label{eq:mamsigma}
m_a^2=\frac {m_u m_d}{(m_u+m_d)^2} \frac {m_\pi^2 f_\pi^2}{f_\varphi^2}
%(n_u+n_d)^2,
\left(\sum_q n_q \right)^2,
\quad\quad m_\sigma^2 = \frac{m_\pi^2 f_\pi^2}{f_\varphi^2}
%\frac{\sum_q (m_q n_q(2-n_q))}{m_u+m_d} 
\frac{\sum_q m_q n_q}{m_u+m_d}
+ 2 \lambda f^2_\varphi, 
\ee
where the former can be obtained from diagonalizing the pseudoscalar mass matrix in the limit $f_\varphi \gg f_\pi$ and is just the standard invisible QCD axion mass. Indeed in this regime  $a$ is just an ordinary QCD axion. One may then use the standard axion mass relation \cite{Weinberg:1977ma}, expressed in terms of the axion decay constant $f_a$, to determine the matching condition
\be 
f_a \equiv \frac{f_\varphi}{\sum_q n_q} \, .
\ee
A light axion is subject to the usual astrophysical constraints. 
Reproducing the desired quark masses needs small Yukawa couplings $\kappa_q$,
\begin{equation}
\label{eq:kumu}
\kappa_u = \sqrt{2} \frac{m_u}{f_\varphi} = 3.1 ~ 10^{-11} \frac{10^{8}\,{\rm GeV}}{f_\varphi}, 
\end{equation}
in the model where $m_u$ only arises from PQ breaking.

The assumption that PQ breaking is triggered by the QCD condensate
has novel cosmological consequences, but also a novel problem.
Reproducing $f_\varphi \gg f_\pi$ from the small QCD-induced PQ breaking demands a small scalar quartic.
So both scalars $a$ and $\sigma$ become extremely light, in a comparable way.
Indeed using eqs.~(\ref{eq:condition-vev})--(\ref{eq:masssigma}) for $m_\varphi^2>0$ one can show,
\begin{equation}
 m^2_a<\frac{m_d}{m_u + m_d} m_\sigma^2 <3  m^2_a ,
\end{equation}
where we again focused on the model where only $m_u$ arises from PQ breaking.

This is problematic due to stringent fifth-force constraints from the exchange of the $\sigma$ scalar. 
The effective scalar-nucleon coupling $g_{\sigma N}$, defined by the interaction term
$\mathscr{L}\supset -g_{\sigma N}\,\sigma_{\rm can}\,\bar N N$,
is given in the isospin limit by \cite{1202.1292}
\begin{equation}
g_{\sigma N}\approx \sum_q \frac{\sigma_q}{m_q}\,y_q ,
\end{equation}
where $y_q$ is the scalar-quark coupling defined through
$\mathscr{L}\supset -y_q\,\sigma_{\rm can}\,\bar q q$, 
and the nucleon sigma terms are approximately 
$\sigma_u\approx 16~\text{MeV}$, $\sigma_d\approx 22~\text{MeV}$, and $\sigma_s\approx 50~\text{MeV}$. Using eq.~\eqref{eq:LQCD},
$y_q = n_q\,m_q/f_\varphi$, we obtain
\begin{equation}
g_{\sigma N}\approx \sum_q n_q\,\frac{\sigma_q}{f_\varphi} 
\approx 1.6 \times 10^{-12} \left( \frac{10^{10} \, \text{GeV}}{f_\varphi} \right),
\end{equation}
where the last benchmark value corresponds again to the 
$\{n_u,n_d,n_s\}=\{1,0,0\}$ model. 

The scalar coupling to nucleons induces a Yukawa-like non-relativistic potential (not spin-suppressed). 
For sufficiently light scalars (in particular in the sub-eV range), this new force can compete with gravity and is therefore tightly constrained by tests of the inverse-square law and searches for equivalence-principle violation (see e.g.~\cite{OHare:2020wah}). 
As a representative benchmark, current limits translate into $g_{\sigma N}\lesssim 10^{-13}$ for $m_\sigma \approx 1~\text{eV}$ (near the edge of astrophysical constraints \cite{2506.19906}), tightening to $g_{\sigma N}\lesssim 10^{-24}$ for $m_\sigma \approx 10^{-16}~\text{eV}$, 
basically ruling out the whole parameter space in the light invisible axion limit.
Similar results are obtained in models with $n_q = 2$, in which the PQ 
symmetry breaking 
is triggered by the QCD quark condensate by turning the 
sign of the $|\varphi|^2$ mass term. Also in this case the radial mode cannot be decoupled from the axion and fifth-force constraints 
exclude this scenario.

Therefore the only viable option appears to add another source of spontaneous PQ breaking (such as $m_\varphi^2<0$), recovering a DFSZ-like scenario. 

\section{Conclusions}

We critically examined the possibility proposed in \cite{Murayama:2026ioh} that the axion might be hidden within hadronic resonances. 
The novel idea is to introduce a complex scalar field charged under a PQ symmetry, whose breaking is triggered by the QCD chiral condensate.
In this way, the up-quark mass term is dynamically generated after confinement, 
reviving the logic of the $m_u = 0$ solution to the strong CP problem. 
If realized in nature, this mechanism would imply that the new scalar degrees of freedom 
are already observed among known hadrons, with the $\eta(1295)$ being a plausible candidate for the pseudo-scalar component. However the model predicts an order unity splitting among charged and neutral pions~\cite{Murayama:2026ioh}. We confirm this problem, fixing order unity factors. Avoiding this problem relying on the Kaplan-Manohar ambiguity 
to set $m_u \ll m_d$ is almost as excluded as using it to more simply set $m_u =0$.

\medskip

Motivated by the hope of avoiding this phenomenological exclusion, 
we considered uncertainties in pion masses and explored several extensions of the minimal setup. Nevertheless, despite some partial success,
we find that this entire class of models is strongly constrained, 
and in practice excluded, by low-energy mesonic observables.
The underlying difficulty can be formulated as a general obstruction. 
If the new scalar degree(s) of freedom are heavy enough to be hidden among hadronic resonances, they can be integrated out, leading to a modified chiral effective theory for the mesons.
The presence of the PQ symmetry introduces a new spurion that explicitly breaks isospin. As a consequence, one novel dominant operator appears in the chiral Lagrangian ($\mathcal{O}_1^{\rm PQ}$ in eq.~\eqref{eq:OPQ1}), 
with a large coefficient such that its effects are comparable to
those of  usual chiral operators arising from quark masses. 
This alternative PQ-symmetric operator is problematic.
At quadratic order in pions,  it breaks the mass degeneracy among them.
At quartic order this operator induces order-one deviations from the well-tested low-energy pion-pion scattering amplitudes.

While one might attempt to cure the pion mass problem by introducing 
additional operators and tuning parameters,  such modifications leave corrections to meson interactions. 
The core issue is therefore structural:  the proposed mechanism requires a non-standard chiral effective action, with new spurions that break isospin. 

\medskip

Finally, we have considered the opposite limit in which the 
PQ breaking from QCD leads to a large axion decay constant $f_a\gg{\rm GeV}$
approaching the usual regime of a light invisible axion. 
To achieve this, the PQ scalar needs to have a small quartic.
As a result its radial mode is almost as light as the axion.
The resulting light scalar mediates a Yukawa-type force between nucleons, 
which is subject to stringent fifth-force constraints. 
These bounds exclude large portions of the parameter space relevant for solving the strong CP problem.

\small

\section*{Note added}
After completion of this work,~\cite{2607.12956} proposed to avoid the light-quark isospin obstruction discussed here by coupling the PQ field $\varphi$ only to the charm quark. More generally, one may contemplate coupling $\varphi$ to a heavy quark $Q=\{c,b,t\}$. 
Although this removes the light-quark PQ spurion at tree level, the field $a$ necessarily couples as $\mathscr{L}\supset -i m_Q a\bar Q\gamma_5 Q/f_\varphi$. 
For $Q=c,b$, this interaction induces the radiative quarkonium decays $J/\psi\to\gamma a$ and $\Upsilon\to\gamma a$, respectively. 
In particular, for $m_a\ll m_{J/\psi}$ the lattice-QCD result 
of~\cite{2502.06721} 
gives $\operatorname{BR}(J/\psi\to\gamma a)\approx 1.8 \, {\rm GeV}^2/f_\varphi^2$. 
The BESIII bound $\operatorname{BR}(J/\psi\to\gamma+\mathrm{inv})<7~10^{-7}$ therefore requires $f_\varphi\gtrsim1.6~{\rm TeV}$~\cite{2003.05594}, incompatible with the $f_\varphi\sim {\rm GeV}$ regime of~\cite{2607.12956}. Analogous bounds from radiative $\Upsilon$ decays exclude the bottom-quark realization~\cite{1007.4646}. 
Moreover, for a charm- or top-quark coupling, electroweak penguin diagrams generate an $s\to d a$ transition. 
In the absence of cancellations involving additional UV interactions, the resulting $K^+\to\pi^+a$ rate is incompatible with the stringent bounds on $K^+\to\pi^++\mathrm{invisible}$~\cite{2110.10698,2507.17286}. 
Thus, while coupling $\varphi$ to a heavy quark removes the tree-level chiral obstruction, it does not yield a viable minimal realization of the QCD-triggered PQ-breaking mechanism.

\footnotesize\medskip

\subsubsection*{Acknowledgements}
LDL is supported by the European Union -- Next Generation EU and
by the Italian Ministry of University and Research (MUR) 
via the PRIN 2022 project n.~2022K4B58X -- AxionOrigins.
The work of MR and AT  is supported by the
Italian Ministry of University and Research (MUR) via the PRIN 2022 project n. 20228WHTYC (CUP:I53C24002320006).
AVT is funded by the European Union, NextGenerationEU, 
National Recovery and Resilience Plan
(mission~4, component~2) 
under the project \textit{MODIPAC: Modular Invariance in Particle Physics and Cosmology} (CUP~C93C24004940006).
AS gratefully acknowledges the absence of financial support. 
MR and AT would like to thank the Galileo Galileo Institute (GGI) for Theoretical Physics for the kind hospitality.

\appendix

\small

\section{Renormalizable models with more PQ scalars}\label{sec:appfail}
%\subsection{Model with 2 scalars}
As an attempt to tune the pion masses, one can consider models with
more than one GeV-scale PQ scalars and thereby more free parameters.
We here consider two scalars, $\varphi_u$ and $\varphi_d$, that
 give mass to the quarks $u$ and $d$ respectively. The Lagrangian for canonically normalized fields reads
\be\label{eq:renorm-models}
\mathscr{L} \supset |\partial \varphi_u|^2+|\partial \varphi_d|^2- m_{\varphi_u}^2|\varphi_u|^2 - m_{\varphi_d}^2|\varphi_d|^2 - ( \kappa_u\varphi_u \overline{u}_L u_R + \kappa_d\varphi_d \overline{d}_L d_R + \text{h.c.}) - V_{\rm PQ}(\varphi_u,\varphi_d)\,.
\ee
The potential $V_{\rm PQ}(\varphi_u,\varphi_d)$  preserves one combination of the two  U(1)'s acting on the two scalars, identified as the PQ symmetry.
%and still it does not contain quartic couplings. \AS{why?} \AT{to make tadpole dominate}
The potential contains only one term of the type 
$(\varphi_u^{n_d}\varphi_d^{n_u}+\text{h.c.})$ or $(\varphi_u^{n_d}{\varphi_d^*}^{n_u}+\text{h.c.})$. These terms effectively realize the following charge assignments $(n_u,n_d)=\{(1,1), (2,\pm1), (\pm 1,2), (3,\pm1), (\pm 1, 3)\}$, if only one of such terms is present at a time. 

Writing the vacuum expectation values as
$\langle \varphi_{u,d} \rangle \equiv f_{u,d}/\sqrt{2}$, the spontaneous breaking of the two U(1)'s together with their diagonal breaking via $V_{\rm PQ}$ allows us to identify the scale $f_\varphi$, appearing in eq.~\eqref{eq:LQCD},  as
\be
\varphi_{u,d}=\frac{f_{u,d}+\sigma_{u,d}}{\sqrt{2}}e^{i a_{u,d}/f_{u,d}}\,,\quad \quad f_\varphi^2 \equiv f_u^2 n_d^2 +f_d^2 n_u^2.
\ee
The $V_{\rm PQ}$ term allows us to identify the new pseudoscalar heavy state $A$, corresponding to the linear combination $n_d a_u/f_u \pm n_u a_d/f_d$, where the $\pm$ refers to terms without and with the complex conjugation. The axion $a$ is the orthogonal combination. 
Diagonalization brings the scalars to the form
\be
\varphi_u = \frac{f_u+\sigma_u}{\sqrt{2}}\exp\bigg( i n_u \frac{a}{f_\varphi}+ i\frac{f_d}{f_\varphi}n_d \frac{A}{f_\varphi}\bigg)\,,\qquad \varphi_d = \frac{f_d+\sigma_d}{\sqrt{2}}\exp\bigg(\pm (i n_d \frac{a}{f_\varphi}- i \frac{f_u n_u}{f_\varphi} \frac{A}{f_\varphi})\bigg) \,,
\ee
where now both $a$ and $A$ have canonically normalized kinetic terms and $\pm$ corresponds to the term with and without the complex conjugation.  This parametrization correctly identifies the charges of the SM quarks and the action of PQ on the Goldstone boson $a$. Indeed, in eq.~\eqref{eq:renorm-models} the axion $a$ now only appears in the Yukawa terms, while $A$ also appears in $V_{\rm PQ}$.  
%In the absence of $V_{\rm PQ}$ the two U(1)'s are spontaneously broken by the chiral condensate with vevs $f_{u,d}\propto \kappa_{u,d} B_0 f_\pi^2/m_{\varphi_{u,d}}^2$ respectively as in \eqref{eq:LQCD}, subject to constraints similar to \eqref{eq:condition-vev}. \AS{cannot understand}
%Now, the presence of $V_{\rm PQ}$ modifies the minimization condition, and it needs to be studied case by case. 
The mass of $A$ is controlled by the size of $V_{\rm PQ}$.
We focus on PQ charges $(n_u, n_d)=(1,1)$, such that  $V_{\rm PQ}$ is simply a mass term:
\begin{equation}
V_{\rm PQ}(\varphi_u,\varphi_d)\equiv\mu_{ud}^2\, \varphi_u \varphi_d^* + \text{h.c.}
\end{equation}
This interaction breaks the rotation of up and down quark to the diagonal combination where the quarks have the same charge. 
In this case, we find
\begin{equation}
 \langle\varphi_u\rangle \equiv \frac{f_u}{\sqrt{2}} 
 = \frac{B_0 f_\pi^2}{2}\, \frac{\kappa_u m_{\varphi_d}^2 - \kappa_d \mu_{ud}^2}{m_{\varphi_u}^2 m_{\varphi_d}^2 - \mu_{ud}^4}
 \qquad \text{and} \qquad
 \langle\varphi_d\rangle \equiv \frac{f_d}{\sqrt{2}}
 = \frac{B_0 f_\pi^2}{2}\, \frac{\kappa_d m_{\varphi_u}^2 - \kappa_u \mu_{ud}^2}{m_{\varphi_u}^2 m_{\varphi_d}^2 - \mu_{ud}^4}\,. 
\end{equation}
Consequently, 
\begin{align}
 m_{\varphi_u}^2 &= \frac{B_0 f_\pi^2 \kappa_u}{\sqrt{2} f_u} - \frac{f_d}{f_u} \mu_{ud}^2 = \frac{B_0 f_\pi^2}{2}\, \frac{\kappa_u^2}{m_u} - \frac{\kappa_u}{\kappa_d}\, 
 \frac{m_d}{m_u}\, \mu_{ud}^2\,, \\
 m_{\varphi_d}^2 &= \frac{B_0 f_\pi^2 \kappa_d}{\sqrt{2} f_d} - \frac{f_u}{f_d} \mu_{ud}^2 = \frac{B_0 f_\pi^2}{2}\, \frac{\kappa_d^2}{m_d} - \frac{\kappa_d}{\kappa_u}\, 
 \frac{m_u}{m_d}\, \mu_{ud}^2\,,
\end{align}
where in the last equalities we have taken into account that
\begin{equation}
 m_u = \frac{\kappa_u f_u}{\sqrt{2}}
 \qquad \text{and} \qquad
 m_d = \frac{\kappa_d f_d}{\sqrt{2}}\,.
\end{equation}
Expanding the scalars in components as
\begin{equation}
 \varphi_{u} = \frac{f_u+\sigma_u}{\sqrt{2}} e^{i \frac{a_u}{f_u}}
 \qquad \text{and} \qquad
 \varphi_{d} = \frac{f_d+\sigma_d}{\sqrt{2}} e^{i \frac{a_d}{f_d}}\,,
 \end{equation}
we see that the $\mu$-term generates a contribution to the potential of 
$a_{u,d}$ of the form
\begin{equation}
 \mu_{ud}^2 f_u f_d \cos\left(\frac{a_u}{f_u} - \frac{a_d}{f_d}\right)\,.
\end{equation}
Adding it to the chiral Lagrangian leads to the following squared mass matrix of pseudo-scalars:
%
%\begin{equation}
% %
% \mathcal{L}_\mathrm{chiral}
% %
%\end{equation}
%
%
\begin{equation}
 %\small
 %
 \mathcal{M}^2 = \begin{pmatrix}
 m_{\varphi_u}^2 & \mu_{ud}^2 & - \frac{B_0 f_\pi \kappa_u}{\sqrt2} & - \frac{B_0 f_\pi \kappa_u}{\sqrt6} & - \frac{B_0 f_\pi \kappa_u}{\sqrt3} \\[0.2cm]
 %\mu_{ud}^2 
 - & m_{\varphi_d}^2 & \frac{B_0 f_\pi \kappa_d}{\sqrt2} & -\frac{B_0 f_\pi \kappa_d}{\sqrt6} & -\frac{B_0 f_\pi \kappa_d}{\sqrt3} \\[0.2cm]
 %- \frac{B f_\pi \kappa_u}{\sqrt2} & \frac{B f_\pi \kappa_d}{\sqrt2} 
 - & - & B_0(m_u+m_d) & \frac{B_0}{\sqrt3}(m_u-m_d) & \sqrt\frac{2}{3} B_0 (m_u-m_d) \\[0.2cm]
 %- \frac{B f_\pi \kappa_u}{\sqrt6} & - \frac{B f_\pi \kappa_d}{\sqrt6} & \frac{B}{\sqrt3}(m_u-m_d) 
 - & - & - &\frac{B_0}{3}(m_u+m_d+4m_s) & \frac{\sqrt2}{3} B_0 (m_u+m_d-2m_s) \\[0.2cm]
 %- \frac{B f_\pi \kappa_u}{\sqrt3} & - \frac{B f_\pi \kappa_d}{\sqrt3} & \sqrt\frac{2}{3} B (m_u-m_d) & \frac{\sqrt2}{3} B(m_u+m_d-2m_s) 
 - & - & - & - & \frac{2}{3} B_0 (m_u+m_d+m_s)+m_1^2
 \end{pmatrix}.
\end{equation}
whose determinant is
\begin{equation}
 \det\mathcal{M}^2 = -\frac{8}{3} B_0^3 f_\pi^2 m_1^2 m_s \mu_{ud}^2  \kappa_u\kappa_d\,.
\end{equation}
If $\kappa_{u,d} > 0$, one needs $\mu_{ud}^2 < 0$ to have all pseudo-scalar masses squared positive. 

We performed a scan of the parameter space of this model, 
but we do not find  values that reproduce  the observed spectrum of $\pi^0$, $\eta$, $\eta'$ while 
all extra scalars have heavy enough GeV masses.
Consistently with the effective theory argument of section~\ref{sec:EFT}, fixing the pion mass problem needs extra lighter scalars, that cannot be hidden
among QCD resonances. A similar issue is found studying more general models.

\section{Effective pion potential}\label{sec:loop}
We here show how loop corrections to the pion mass splitting can be computed in the perturbative regime. This also provides an estimate of the splitting in the relevant strong coupling regime discussed in the paper.  

The computation is analogous to the electromagnetic splitting of pions using form factors. Following \cite{1005.4269}, we can compute the effective potential of pions and extract the $\pi^\pm-\pi^0$ splitting at 1-loop. In the limit where the couplings $\kappa_q$ are treated as small perturbations, the effective action for $\varphi$ in the background of $\pi$'s can be computed by writing down the most general action consistent with the symmetry of the chiral Lagrangian. To do so we upgrade $\kappa_q\varphi$ to a field $\Phi$ transforming $\Phi \to L \Phi R^\dag$ under the chiral symmetry. Treating $\Phi$ as an external source to quadratic order and in the chiral limit, the effective action in the background of pions is 
\be
\mathscr{L}_{\rm eff}=\Pi_A(p) \tr[\Phi \Phi^\dag] + \Pi_B(p)\tr[\Phi U^\dag] \tr[\Phi^\dag U]+ \Pi_0 \tr[\Phi U^\dag + \text{h.c.}]\,.
\label{eq:effaction}
\ee
The form factors $\Pi_{A,B}(p)$ are determined by the two point function of the QCD operators $\Op_{LR}^{ij}\equiv\bar q_L^i q_R^j$. Exploiting the SU($N_F$) flavour symmetry and parity we decompose $\Op_{LR}^{ij}$ into irreps as,
 \be\label{eq:singlet-octet}
\Op_{LR}^{ij}= \frac{\delta^{ij}}{\sqrt{2N_F}} \Op_{\bf 1} + T^a_{ij}\Op_{\bf Adj}^a\,,\qquad \Op_{\mathbf 1} = (\Op_{\mathbf{1},S} - i\Op_{\mathbf{1},P})\,, \qquad \Op_{\bf Adj}^a = (\Op_{\mathbf{{\bf Adj}},S}^a -i \Op_{\mathbf{{\bf Adj}},P}^a)\,,% 
 \ee 
where $T^a$ are the generators in the adjoint of SU($N_F$). For the relevant case $N_F=3$ but we write formulas for general number of flavors. The iso-spin channels are the singlet and the octet with complex operators $\Op_{(\mathbf{1},\mathbf{{\bf Adj}})}$. In eq.~\eqref{eq:singlet-octet} they have been written as the sum of CP-even (S) and CP-odd (P) operators. Explicitly the $S$ operators are $\Op_{\mathbf{1}, S}\equiv \sqrt{2/N_F}\bar q q$ and the $P$ are $\Op_{\mathbf{1}, P}\equiv \sqrt{2/N_F} i\bar q \gamma_5 q$, and similarly for the channel in the adjoint. 
Such operators interpolate with QCD states with the same symmetry properties.
We are interested in the momentum space transform of correlators $\langle \Op(x)\Op(0)\rangle$, given by 
\be
\langle \Op_{\mathbf{1},(S,P)}(p)
\Op_{\mathbf{1},(S,P)}(-p)\rangle=i\Pi_{\mathbf{1},(S,P)}(p)\,, \quad
 \langle \Op_{\mathbf{{\bf Adj}},(S,P)}^a(p)\Op_{\mathbf{{\bf Adj}},(S,P)}^b(-p)\rangle=i\Pi_{\mathbf{{\bf Adj}},(S,P)}(p)\delta^{ab}\,.
\label{eq:eff}
%\end{split}
\ee
These  correlators only depend on the QCD dynamics and are extracted in the chiral limit. While the general form of the 2-point functions depends on the details of QCD, the situation simplifies in the large $N$ limit where they become a sum of single poles. Moreover the contribution of Nambu-Goldstone bosons is determined by the chiral Lagrangian. We thus have,
\begin{equation}\label{eq:form-factors}
\Pi_{\mathbf{1},P} = \frac{f_{\eta'}^2 B_0^2}{p^2-m
_{\eta'}^2} +\cdots\,,\quad\quad
\Pi_{\mathbf{{\bf Adj}},P} = \frac{f_{\pi}^2 B_0^2}{p^2-m_\pi^2} +\cdots\,,\quad
\Pi_{\mathbf{1},S} = \frac{f_{\sigma_0}^2 B_0^2}{p^2-m
_{\sigma_0}^2} +\cdots\,,\quad\quad
\Pi_{\mathbf{{\bf Adj}},S} =\frac{f_{\sigma_8}^2 B_0^2}{p^2-m_{\sigma_8}^2} +\cdots.
\end{equation}
In real QCD  $f_{\eta'}\approx 1.3 f_\pi$ and the lightest scalar resonances can be identified with $f_0(500)$ and isospin triplet $a_0(980)$ whose decay constants are not known but are expected to be of order $f_\pi$.
The form of $\Pi'$s is easily understood: the leading contribution to each form factor arises from the exchange of QCD single-particle states with the same quantum numbers of the source, namely, singlet $\eta'$, meson octet, singlet $\sigma$ resonance and octet of CP-even scalars. The dots represent contributions from heavier resonances and multiparticle states. 

The form factors $\Pi_{A,B}$ in eq. (\ref{eq:effaction}) are determined by QCD correlation functions above. This can be derived by matching $\Pi_{\bf 1}$ and $\Pi_{\bf Adj}$ to the Lagrangian in eq.~\eqref{eq:effaction} on the vacuum $U=1$. Parametrizing $\Phi = \Phi_1/\sqrt{2N_F} + \Phi_{\bf Adj}^a T^a$ with $\tr[T^aT^b]=\delta^{ab}/2$, we obtain,
\be
\Pi_A = -\frac{ \Pi_{\mathbf{{\bf Adj}},S} +\Pi_{\mathbf{{\bf Adj}},P}}{2}\,,\qquad 
\Pi_B=\frac{ \Pi_{\mathbf{{\bf Adj}},S} + \Pi_{\mathbf{{\bf Adj}},P}-\Pi_{\mathbf{1},S} -\Pi_{\mathbf{1},P}}{2N_F}\,.
\ee
This is reminiscent of the $V-A$ structure of the electromagnetic splitting \cite{1005.4269}. From the chiral Lagrangian we also extract $\Pi_0= f_\pi^2 B_0/2$.

Assuming $\Pi_{A,B}$ are known we can proceed to the calculation of the effective potential generated by $\varphi$. For this we make  $\varphi$ dynamic adding its action, $|\partial_\mu \varphi|^2-m_\varphi^2 |\varphi|^2$, to eq. (\ref{eq:eff}). For simplicity we can work with only pions in the  matrix $U$ so that exponential of the matrix can be explicitly written down. 
We find,
\be
\begin{split}
\mathscr{L}_{\rm eff}&=
\varphi(p)\varphi^*(-p)\bigg\{ (p^2-m_\varphi^2)  + \Pi_A(p) (\kappa_u^2 n_u^2+\kappa_d^2 n_d^2) \\
&+\Pi_B(p)\left[(\kappa_u n_u+\kappa_d n_d)^2 \cos^2(\pi)+ (\kappa_u n_u-\kappa_d n_d)^2\frac{(\pi^0)^2}{\pi^2}\sin^2(\pi)\right]\bigg\}\\
&+\Pi_0 \varphi(p) \left[(\kappa_u n_u+\kappa_d n_d) \cos(\pi) +i (\kappa_u n_u-\kappa_d n_d)\frac{\pi^0}{\pi}\sin(\pi)\right]+\text{h.c.}
\end{split}
\ee
With these expressions we can easily determine the effective potential performing the path integral over $\varphi$. The term linear in $\varphi$ on the third line generates the tree level contribution obtained integrating out $\varphi$ in section \ref{sec:EFT}. Quantum fluctuations produce a correction, 
\be
\begin{split}
\Delta V_{\rm eff}(\pi)=  &2\times \frac{-i}{2}\int \frac{d^4p}{(2\pi)^4}\ln\left[p^2-m_\varphi^2+\left(\sum \kappa_q^2 n_q^2\right)\Pi_A(p)+\right.\\
&\left. +\Pi_B(p)\left((\kappa_u n_u+\kappa_d n_d)^2 \cos^2\pi+ (\kappa_u n_u-\kappa_d n_d)^2\frac{(\pi^0)^2}{\pi^2}\sin^2\pi\right)\right]\,.
\end{split}
\ee
Expanding the logarithm, and Wick rotating to Euclidean momenta $p^2\to -Q^2$, at leading order in $\kappa_{u,d}$ we can extract the contribution to the pion splitting,
\be
\Delta(m_{\pi^0}^2-m_{\pi^\pm}^2)\approx (\kappa_u n_u-\kappa_d n_d)^2\times\frac 2 {f_\pi^2}\int \frac{d^4Q}{(2\pi)^4}\frac{\Pi_B(Q^2)}{-Q^2-m_\varphi^2}\,.
\ee

Knowing the full momentum dependence of the form factors one can precisely determine the mass splitting.  Similarly to the electromagnetic splitting we can estimate the integral assuming that it is dominated by the lowest resonances. In the limit where $m_\varphi$ is heavy we can require that the integral is finite fixing decay constants $f_{\sigma_{0,8}}$.
We find,
\begin{equation}
\int \frac{d^4Q}{(2\pi)^4}\Pi_B(Q^2)= -\frac{
B_0^{2} \left[
f_{\eta'}^{2} m_{\eta'}^{2} (m_{a_0}^{2} - m_{\sigma}^{2})
\ln\!\left(\frac{m_{\eta'}^{2}}{m_{a_0}^{2}}\right)
+
\left( f_{\pi}^{2} m_{a_0}^{2}
+ f_{\eta'}^{2} (m_{\eta'}^{2} - m_{a_0}^{2}) \right)
m_{\sigma}^{2}
\ln\!\left(\frac{m_{\sigma}^{2}}{m_{a_0}^{2}}\right)
\right]
}
{32\pi^2 \,  (m_{a_0}^{2} - m_{\sigma}^{2})\, \, N_F}.
\end{equation}
Numerically using the observed values of the masses and $N_F=3$ we find that the correction to the mass of $\pi_0$ is negative and smaller than the tree-level contributions. However, there is no parametric suppression in the result so that the contribution remains uncertain especially in the relevant regime $\kappa_q\sim 1$.

\end{document}